# Hyperuniform disordered photonic crystal polarizers

Wen Zhou,[a)] Yeyu Tong, Xiankai Sun, and Hon Ki Tsang

*Department of Electronic Engineering, The Chinese University of Hong Kong, Shatin, New Territories, Hong Kong*

An ultra-broadband transverse magnetic (TM) pass hyperuniform disordered photonic crystal (HUDPC) polarizer is proposed and demonstrated on a silicon-on-insulator platform. Propagation of the transverse electric mode is blocked by three combined effects, including the photonic bandgap (PBG) effect, diffusive (non-resonant) scattering, and bandedge resonances. Specially, the designed 30-dB bandwidth in polarization extinction ration (PER) of 265 nm is much larger than the spectral width of the PBG (149 nm) due to using the bandedge resonances. The TM mode is in the subwavelength regime of the HUDPC and thus has a low insertion loss (IL). An ultrawide 30-dB bandwidth in PER of 210 nm (1.44–1.65 µm) is experimentally demonstrated in a 12.9-µm-long HUDPC polarizer with spectrally averaged PER of 39.6 dB and IL for the TM mode of 1.1 dB (IL = 0.6 dB at 1.55 µm). The HUDPC polarizers can be an excellent candidate for ultra-broadband polarization filtering in the silicon photonic platform.

## I. INTRODUCTION

The generalization of the hyperuniform point configurations to the heterogeneous materials for the disordered photonics[1] is an analog of the hyperuniform configuration of particles in the condensed matter physics.[2,3] The sufficient condition for a hyperuniform point pattern is to have the point positions that can produce structure factor which tends to be zero as the wave vector tends to zero,[2] while the stealthy-type hyperuniform patterns have structure factors that are zero for a subset of their wave vectors around the origin.[3] Without satisfying the Bloch's theorem, the isotropic, sizeable, and even complete photonic bandgaps (PBGs) were discovered in a stealthy-type HUDPC and its derivative wall network with low-density fluctuation and short-range order.[4,5] Different types of photon transport can exist in a HUDPC, including transparency,[6] diffusion, localization, and propagation prohibition based on the PBG effect.[7] With these distinct properties, HUDPCs and hyperuniform wall networks have found applications in freeform microwave waveguides,[5] hollow-core terahertz waveguides,[8] microwave Luneburg lens,[9] photonic spectrometers,[10] and photonic network lasers.[11]

In this work, we show that a HUDPC-patterned waveguide can be designed to serve as an ultra-broadband transverse magnetic (TM) pass/transverse electric (TE) block polarizer on a silicon-on-insulator (SOI) platform. Previous approaches for the SOI waveguide polarizers are based on the resonant tunneling in the asymmetric waveguide coupler structure,[12] hybrid plasmonic effect in a highly doped silicon waveguide,[13] mode leakage in the tight waveguide bends,[14] mode cut-off in the ridge waveguides,[15] and subwavelength grating (SWG) waveguides,[16] polarization-dependent photonic bandgap (PBG) effect in the 2D photonic crystal (PhC) waveguides,[17] SWG

---

[a)] Author to whom correspondence should be addressed: wzhou@ee.cuhk.edu.hk

waveguides,[18,19] and hyperuniform disordered wall networks.[20] Specifically, a polarization extinction ratio (PER) of 30 dB was demonstrated in a 15-μm-long hybrid plasmonic waveguide at the cost of 3-dB insertion loss (IL).[13] A 30-dB bandwidth in PER of 100 nm was demonstrated in the cascaded waveguide bends with a footprint of 63 μm × 9.5 μm.[14] Mode cut-off was introduced in a periodic SWG with a designed 30-dB bandwidth exceeding 200 nm and a PER of 35 dB, however, a long device length of 60 μm is required.[16] With the polarization-dependent PBG effect, the PER was enhanced to 40 dB in a 17.6-μm-long SWG waveguide with a 0.4-dB IL measured at a single wavelength of 1.55 μm, while the 30-dB bandwidth was not reported.[18] A hyperuniform disordered wall network polarizer was demonstrated based on the PBG effect and diffusive (non-resonant) scattering to block the TE mode, with a 30-dB bandwidth of 98 nm and an averaged IL of 1.7 dB.[20]

The proposed HUDPC polarizers are designed on a 220-nm SOI wafer with a strong birefringence. The 30-dB bandwidth of the HUDPC polarizer can be significantly improved to 265 nm while the IL is as low as 1.1 dB due to weak confinement of the TM mode in the HUDPC, compared with those (30-dB bandwidth = 98 nm and IL = 1.7 dB) of the hyperuniform disordered wall network polarizers designed on the 340-nm SOI.[20] By analyzing the reflection and scattering spectra, the proposed HUDPC polarizer operates with a new principle, i.e., transmission of the TE-polarized light is blocked by three combined effects, including the PBG effect, diffusive (non-resonant) scattering, and bandage resonances, while the TM-polarized light operates in the subwavelength regime.[21,22] Specially, with presence of the bandedge resonances,[11] the designed 30-dB bandwidth (265 nm) is much larger than the spectral width of the PBG (149 nm) in our proposed HUDPC polarizer, which bypasses the bandwidth limitation that is smaller than the size of PBG in the conventional PBG-based polarizers.[18,20] Meanwhile, the spectrally averaged back reflection is less than 60%, which is much lower than those of the PBG-based polarizers[18,20] due to resonantly enhanced scattering using the bandedge resonances. By increasing the device length to 21.5 μm, the combined effects in the HUDPC polarizer support a designed 30-dB bandwidth of 296 nm with spectrally averaged IL and PER of 1 dB and 38 dB, respectively. The experimental results show a 30-dB bandwidth exceeding 210 nm (1.44–1.65 μm), a spectrally averaged PER of 39.6 dB, and an averaged IL for the TM mode of 1.1 dB (IL = 0.6 dB at 1.55 μm) in a 12.9-μm-long HUDPC polarizer. Compared with those of the previously demonstrated polarizers,[14,16,18,19,20] our demonstrated 30-dB bandwidth and PER are at the state-of-the-art level with a relatively short device length and a low IL.

## II. DESIGN AND WORKING PRINCIPLE

Figure 1(a) shows a schematic of the proposed TM-pass/TE-block HUDPC polarizer. The polarizer is a SOI channel waveguide patterned by disordered air holes based on a stealthy-type point pattern with a constraint factor $\chi$ of 0.5, produced using the collective coordinate approach.[3,23] Figure 1(b) shows top view and labelled parameters. The diameters, local periods of air holes, and total length of a HUDPC are denoted by $D_{hole}$, $\Lambda_{hole}$, and $L_{device}$, respectively. Figure 1(c) shows cross section of a SOI channel waveguide with labelled width of $W_{WG}$ and height of $H_{WG}$. The upper and under claddings are PMMA and buried oxide (BOX) with heights of $H_1$ and $H_2$.

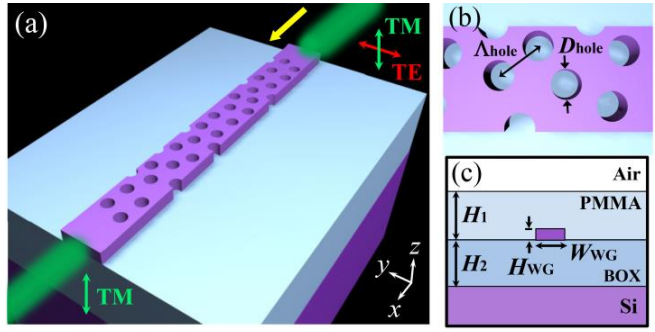

FIG. 1. Schematic of a TM-pass/TE-block hyperuniform disordered photonic crystal (HUDPC) polarizer. (a) Angled overview. (b) Top view and zoom in on disordered air holes with labelled parameters. (c) Cross section of a waveguide with labelled materials and parameters.

The TE-polarized light is strongly reflected due to the PBG effect when a HUDPC is designed with $\langle \Lambda_{hole} \rangle \sim \lambda/(2n_{eff\_TE})$, where $\langle \Lambda_{hole} \rangle$, $\lambda$, and $n_{eff\_TE(TM)}$ are the average period, working wavelength, and effective refractive index (RI) of the TE(TM) mode, respectively. There is non-resonant scattering loss when light diffuses in a disordered lattice. Especially, scattering can be resonantly enhanced at two bandedges of the PBG of a HUDPC. A combination of the PBG effect, diffusive scattering, and bandedge resonances contributes to blocking of the TE-mode. Due to a large birefringence ($n_{eff\_TE} > n_{eff\_TM}$), the subwavelength condition $\langle \Lambda_{hole} \rangle < \lambda/(2n_{eff\_TM})$ is simultaneously satisfied by the TM mode,[24,25] which passes through a HUDPC with little scattering and reflection loss.

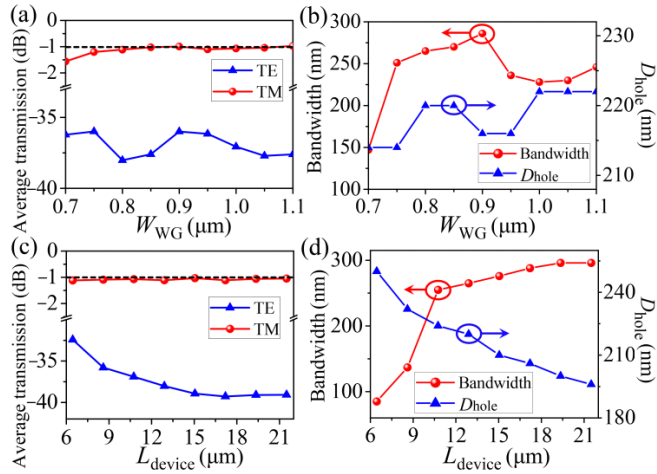

FIG. 2. Numerical investigation. (a, b) Performance as a function of $W_{WG}$ and $D_{hole}$ in average transmission (a) and 30-dB bandwidth (b) when $L_{device}$ = 12.9 μm. (c, d) Performance as a function of $L_{device}$ and $D_{hole}$ in average transmission (c) and 30-dB bandwidth (d) when $W_{WG}$ = 800 nm.

In following, ($H_1$, $H_2$, $H_{WG}$, $\langle \Lambda_{hole} \rangle$) are fixed as (1.4, 3.0, 0.22, 0.378) μm. The performance of a HUDPC polarizer was numerically investigated based on a 3D finite-difference time-domain (FDTD) method[26] with a grid size of 10 nm. An etch angle of 2° was incorporated into the simulation model according to our scanning electron microscopic (SEM) characterization. The fundamental TE/TM mode was launched into a bus waveguide. After a HUDPC polarizer, the total transmission was recorded for the TE mode [$T_{TE}(\lambda)$], while an eigenmode expansion monitor was used to extract the transmission of the fundamental TM mode [$T_{TM}(\lambda)$]. We defined a 30-dB bandwidth as to simultaneously satisfy the polarization extinction ratio PER($\lambda$) ≥ 30 dB and insertion loss of the fundamental TM mode $IL_{TM}(\lambda) \leq 3$ dB, where PER($\lambda$) = $10\log_{10}(T_{TM}(\lambda)/T_{TE}(\lambda))$, $IL_{TM(TE)}(\lambda) = -10\log_{10}(T_{TM(TE)}(\lambda))$ (unit: dB). Figures 2(a) and 2(b) show performance variation of a HUDPC polarizer ($L_{device}$ = 12.9 μm) by scanning $W_{WG}$ from 0.7 to 1.1 μm, while the spectrally averaged transmission of the fundamental TM mode $\langle T_{TM}(\lambda) \rangle$ is maintained at around −1.0 dB by adjusting $D_{hole}$ for each $W_{WG}$. 30-dB bandwidths are 288 nm ($\lambda$ = 1.431–1.503 and 1.522–1.736 μm), 271 nm ($\lambda$ = 1.413–1.425 and 1.463–1.72 μm), and 265 nm ($\lambda$ = 1.443–1.707 μm) when ($W_{WG}$, $D_{hole}$) are (0.9, 0.216), (0.85, 0.22), and (0.8, 0.22) μm, respectively. We choose ($W_{WG}$, $D_{hole}$) of (0.8, 0.22) μm to obtain a continuous and large 30-dB bandwidth. Figures 2(c) and 2(d) show that $\langle IL_{TE}(\lambda) \rangle$ and 30-dB bandwidth of a HUDPC polarizer ($W_{WG}$ = 0.8 μm) increase simultaneously by scanning $L_{device}$ from 6.5 to 21.5 μm, while $\langle T_{TM}(\lambda) \rangle$ is maintained at around −1.0 dB by adjusting $D_{hole}$ for each $L_{device}$. There is a turning point at ($L_{device}$, $D_{hole}$) = (10.8, 0.224) μm with a 30-dB bandwidth of 255 nm ($\lambda$ = 1.447–1.701 μm), after which 30-dB bandwidth further increases to 296 nm ($\lambda$ = 1.457–1.752 μm) with ($L_{device}$, $D_{hole}$) = (21.5, 0.196) μm.

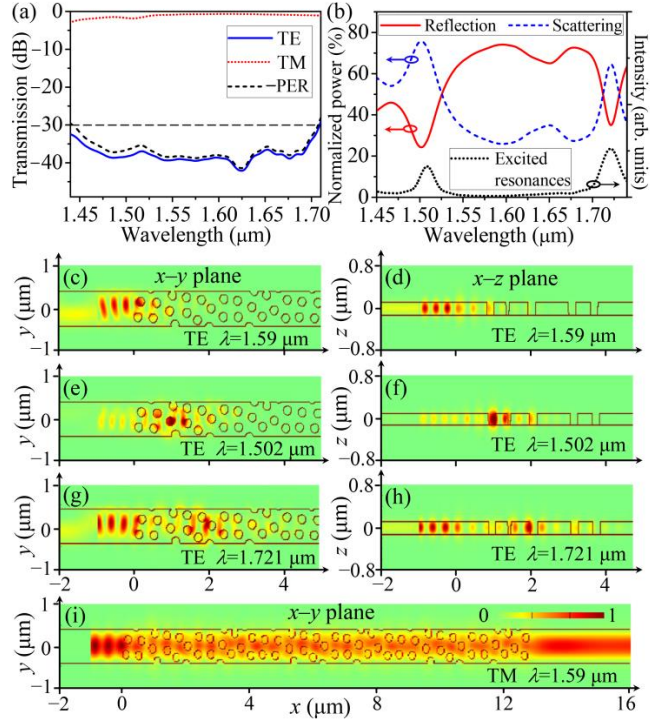

FIG. 3. Performances of a HUDPC polarizer with ($L_{device}$, $W_{WG}$, $\langle \Lambda_{hole} \rangle$, $D_{hole}$) = (12.9, 0.8, 0.378, 0.22) μm. (a) Normalized spectra in transmission of the TE (TM) mode and −PER. (b) Normalized spectra in back reflection and scattering of the TE mode and intensity spectrum of excited bandedge resonances. (c)–(h) $|E|^2$ profiles of the TE-polarized light at $\lambda$ = 1.59 μm in top (c) and side views (d), at $\lambda$ = 1.502 μm in top (e) and side views (f), and at $\lambda$ = 1.721 μm in top (g) and side views (h). (i) Top-view $|E|^2$ profile of the TM-polarized light at $\lambda$ = 1.59 μm.

Figure 3(a) shows simulated transmission spectrum of the TE(TM) mode and the corresponding −PER spectrum for a HUDPC polarizer with ($L_{device}$, $W_{WG}$, $\langle\Lambda_{hole}\rangle$, $D_{hole}$) = (12.9, 0.8, 0.378, 0.22) μm. The spectrally averaged transmission of the TE and TM modes [$\langle T_{TE}(\lambda)\rangle$ and $\langle T_{TM}(\lambda)\rangle$] are respectively −38.0 and −1.1 dB within a 30-dB bandwidth of 265 nm (1.443–1.707 μm) when $IL_{TM}(\lambda) \leq 3$ dB and $PER(\lambda) \geq 30$ dB, simultaneously. Figure 3(b) shows the back-reflection (red solid curve) and scattering spectra (blue dashed curve) for the blocked TE mode. Inside the 30-dB bandwidth, the spectrally averaged scattering and back reflection are respectively 41.0% and 59.0%, which is much smaller than those of the periodic SWG-based polarizers (95%)[18] and that of a hyperuniform disordered wall network polarizer (82%).[20] Especially, the back reflection is only 24% at 1.502 μm. The back reflection becomes dominant in the wavelength range of 1.525–1.712 μm mainly due to the PBG effect. We adopted a dipole cloud approach[27] to calculate the intensity spectrum of excited resonances (black dotted curve) for the HUDPC polarizer as shown in Fig. 3(b). The PBG is defined by a spectral region, in which the intensity of excited resonant modes [$I(\lambda)$] are suppressed by the PBG effect and are smaller than 10% of the maximal intensity [$I(\lambda) \leq 0.1 \cdot \max(I)$] in the intensity spectrum. The calculated wavelength range for the PBG is between 1.532–1.681 μm (149 nm) of the HUDPC polarizer with ($L_{device}$, $W_{WG}$, $\langle\Lambda_{hole}\rangle$, $D_{hole}$) = (12.9, 0.8, 0.378, 0.22) μm. Scattering dominates in the wavelength ranges of 1.443–1.525 μm and 1.712–1.731 μm, because of excitation of bandedge radiative resonances, which are spatially localized and have low $Q$ factors.[11] To validate, the calculated excited resonances using a dipole cloud approach[27] are at 1.508 and 1.721 μm as shown in Fig. 3(b).

The 3D FDTD simulations produce the $|E|^2$ profiles which are plotted in Figs. 3(c)–3(i). The TE-polarized light at 1.59 μm experiences the PBG effect and quickly attenuates from the beginning of a HUDPC ($x = 0$ μm) with strong back reflection (without localization inside lattice) as shown in Figs. 3(c) and 3(d). As shown in Figs. 3(e)–3(h), at two scattering peaks ($\lambda = 1.502$ and 1.721 μm in Fig. 3(b)), two bandedge modes are localized at $x = 1.1$ μm and $x = 1.9$ μm in a HUDPC. Bandedge modes at 1.502 μm (1.721 μm) is predominantly concentrated in the air (Si), indicating the air (dielectric) bandedge mode. The PBG is associated with a wavelength region between the dielectric and air bandedge modes,[4,5] in which light is mainly reflected. These random resonances with low $Q$ factors are radiative, resonantly enhancing scattering at two peaks as shown in Fig. 3(b). The TM-polarized light at $\lambda = 1.59$ μm passes through the entire HUDPC ($L_{device} = 12.9$ μm) with a transmission of 85.2% as shown in Fig. 3(i).

## III. FABRICATION AND CHARACTERIZATION

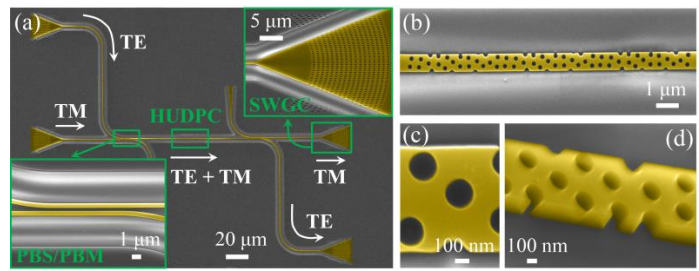

FIG. 4. SEM images. (a) Top view of an entire device under test. Insets show a polarization beam splitter/multiplexer (PBS/PBM) and a subwavelength grating coupler (SWGC) indicated by the green boxes in (a). (b) Top view of an entire HUDPC polarizer ($L_{device} = 12.9$ μm). (c, d) Zoom-in of air holes in top (c) and angled views (d).

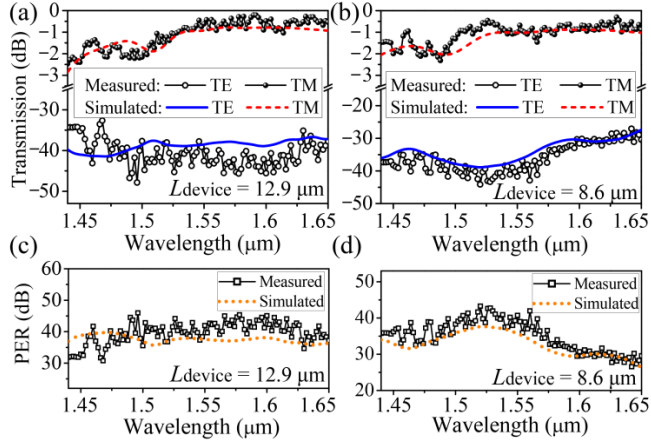

FIG. 5. Experimental demonstration and numerical validation. (a, b) Normalized and stitched transmission spectra for the TE and TM modes in HUDPC polarizers with $L_{device}$ of 12.9 μm (a) and 8.6 μm (b). (c, d) The corresponding PER spectra for HUDPC polarizers with $L_{device}$ of 12.9 μm (c) and 8.6 μm (d).

Devices were fabricated on a 220-nm-thick SOI with a single step of electron-beam lithography and reactive-ion etching. The fabricated devices were characterized with a SEM before spin-coating PMMA for measuring the transmission spectra. Figure 4(a) shows a SEM image of a device under test, including the polarization-dependent focusing subwavelength grating couplers (SWGCs) for launching the TE and TM polarizations into the SOI waveguides.[28,29] The symmetric directional coupler functions as a polarization beam splitter/multiplexer (PBS/PBM) with a designed center wavelength at 1.563 μm and a 3-dB bandwidth over 1.44–1.68 μm. Thus, transmission spectra of the orthogonal polarizations can be measured in a single device. Figures 4(b)–4(d) show top view and tilted view of a HUDPC polarizer with full-etched air holes and smooth sidewalls. The etch angle is estimated to be 2°. The experimental setup can be found in Ref. [20]. The output spectrum is recorded by scanning wavelength in a range of 1.44–1.65 μm. Spectral normalization is given by difference in transmission of devices with and without a HUDPC, to exclude the ILs from SWGCs and PBS/PBM. Due to limited bandwidth of the SWGCs, a single device is tested at several different diffraction angles ($\theta_{diff}$). The optical bandwidths with efficient light coupling are stitched to obtain a broadband spectrum.[30] As shown in Fig. 5(a), for a HUDPC with ($D_{hole}$, $W_{WG}$, $L_{device}$) = (0.21, 0.76, 12.9) μm, measured $\langle T_{TE}(\lambda) \rangle$ and $\langle T_{TM}(\lambda) \rangle$ are respectively −40.7 and −1.1 dB within a 30-dB bandwidth of 1.44–1.65 μm when PER($\lambda$) ≥ 30 dB and $IL_{TM}(\lambda)$ ≤ 3 dB, simultaneously. Measured structural parameters were incorporated into a 3D FDTD model. The predicted 30-dB bandwidth is 257 nm (1.434–1.69 μm), $\langle T_{TE}(\lambda) \rangle$ and $\langle T_{TM}(\lambda) \rangle$ are respectively −39.3 and −1.2 dB. Around $\lambda$ = 1.45 μm, drop in $T_{TM}(\lambda)$ is observed because of the non-negligible diffraction effect of the TM mode at a shorter wavelength. In Fig. 5(b), for a HUDPC with ($D_{hole}$, $W_{WG}$, $L_{device}$) = (0.24, 0.77, 8.6) μm, measured $\langle T_{TE}(\lambda) \rangle$ and $\langle T_{TM}(\lambda) \rangle$ are respectively −38.1 and −1.2 dB within a 30-dB bandwidth of 157 nm (1.44–1.596 μm). Simulated $\langle T_{TE}(\lambda) \rangle$ and $\langle T_{TM}(\lambda) \rangle$ are respectively −36.1 and −1.4 dB inside a 30-dB bandwidth of 168 nm (1.417–1.584 μm). Figures 5(c) and 5(d) show the corresponding PER spectra.

Table 1. Performance of the demonstrated waveguide polarizers

| Ref. | 30-dB bandwidth (nm) | | PER (dB) | | IL (dB) | | $L_{device}$ (μm) | Working principle |
|---|---|---|---|---|---|---|---|---|
| | Sim. | Exp. | Sim. | Exp. | Sim. | Exp. | | |
| 14 | >100 | >100 | 40.7 | 36 | 0.09 | 0.12 | 63 | mode leakage |
| 16 | 200 | 110 | 35 | 30 | 0.3 | 0.4 | 60 | mode cut-off |
| 18 | 80 | 60 (PER ≥ 20) | 34 | 27 | <0.3 | 0.5 | 12.8 | PBG |
| 19 | 95 | 110 (PER ≥ 20) | >30 | 18–29 | 0.1 | 1.4–2.3 | 144 | PBG |
| 20 | 110 | 98 | 32 | 32 | 1.78 | 1.72 | 8 | PBG + diffusive scattering |
| This work | 257 | >210 | 38.1 | 39.6 | 1.2 | 1.1(0.6) | 12.9 | PBG + diffusive scattering + bandedge resonances |

## IV. CONCLUSION

In conclusion, we demonstrated a 30-dB bandwidth of 210 nm (limited by scan range of the tunable laser) with spectrally averaged PER of 39.6 dB and IL of 1.1 dB (IL = 0.6 dB at $\lambda = 1.55$ μm) in a 12.9-μm-long HUDPC polarizer, which show one of the widest bandwidths and highest PERs with a relatively low IL and a short device length as summarized in Table 1. A working principle, which is based on a combination of the PBG effect, diffusive scattering, and bandedge resonances in the HUDPC, was proposed to effectively block transmission of the TE mode. While the TM mode operates in the subwavelength regime with a low insertion loss. Using the bandedge resonances, the designed 30-dB bandwidth in PER (265 nm) is much larger than the spectral width of the PBG (149 nm) in our proposed HUDPC polarizer, which bypasses the bandwidth limitation that is smaller than the size of PBG in the conventional PBG-based polarizers.[18,20] Meanwhile, the spectrally averaged back reflection is less than 60%, which is much lower than those of the PBG-based polarizers[18,20] due to resonantly enhanced scattering using the bandedge resonances. With an excellent and balanced performance, our proposed HUDPS polarizer is promising for the ultra-broadband polarization filtering on a silicon photonic platform.

**ACKNOWLEDGMENTS**

This work was fully supported by Hong Kong Research Grants Council, General Research Fund (RGC, GRF) under Grant 14205515. Wen Zhou acknowledges the support from the Postdoctoral Hub–Innovation and Technology Fund (PH–ITF).

28  Z. Cheng, X. Chen, C. Wong, K. Xu, and H. K. Tsang, Appl. Phys. Lett. **101**, 101104 (2012).
29  W. Zhou, Z. Cheng, X. Sun, and H. K. Tsang, Opt. Lett. **43** (12), 2985 (2018).
30  H. Xu, D. Dai, and Y. Shi, Laser Photon. Rev. **13** (4), 1800349 (2019).